\documentclass[final]{svjour2}
\usepackage{graphicx}
\usepackage{rotating}
\usepackage{amssymb}
\usepackage{mathptmx}
\usepackage[numbers,sort]{natbib}
\makeatletter \journalname{Journal of Low Temperature Physics}
\bibpunct{}{}{,}{s}{}{,}

\begin{document}

\newcommand{\hdblarrow}{H\makebox[0.9ex][l]{$\downdownarrows$}-}
\title{Is a gas of strongly interacting atomic fermions a nearly perfect fluid?}

\author{A. Turlapov$^{1,2}$ \and J. Kinast$^1$ \and B. Clancy$^1$\and \\Le Luo$^1$ \and J. Joseph$^1$ \and J. E. Thomas$^{1,3}$}

\institute{1:Department of Physics, Duke University, Durham, NC
27708-0305, USA
\\2: Current address: Institute of Applied Physics, Russian Academy of Sciences, 46 ul. Ulyanova, Nizhniy Novgorod, 603950, Russia
\\3: \email{jet@phy.duke.edu}}

\date{XX.XX.2007}

\maketitle


\begin{abstract}
We use all-optical methods to produce a highly-degenerate Fermi
gas of spin-$1/2$ $^6$Li atoms. A magnetic field tunes the gas
near a collisional (Feshbach) resonance, producing strong
interactions between spin-up and spin-down atoms. This atomic gas
is a paradigm for strong interactions in nature, and provides
tests of current quantum many-body calculational methods for
diverse systems, including very high temperature superconductors,
nuclear matter in neutron stars, and the quark-gluon plasma of the
Big Bang. We have measured properties of a breathing mode over a
wide range of temperatures.  At temperatures both below and well
above the superfluid transition, the frequency of the mode is
nearly constant and very close to the hydrodynamic value. However,
explaining both the frequency and the damping rate in the normal
collisional regime has not been achieved. Our measurements of the
damping rate as a function of the energy of the gas are used to
estimate an upper bound on the viscosity. Using our new
measurements of the entropy of the gas, we estimate the ratio of
the shear viscosity to the entropy density, and compare the result
with the lower bound for quantum viscosity recently predicted
using string theory methods.

\end{abstract}

\section{Introduction}

Optically-trapped, strongly-interacting atomic Fermi
gases~\cite{OHaraScience,AmScientist} provide a unique laboratory
for testing nonperturbative many-body theories in a variety of
fields, from neutron stars and nuclear
matter~\cite{Heiselberg,Baker,Carlson} to quark-gluon
plasmas~\cite{Heinz} and high temperature
superconductors~\cite{Levin}. In contrast to other Fermi systems,
in atomic gases, one may tune at will the
interactions~\cite{Grimmbeta,GrimmLeeHuangYang,JosephSound},
energy~\cite{JointScience,KinastDampTemp}, and spin
populations~\cite{KetterleImbalanced,HuletPhaseSeparation}. The
strong interactions are produced using a Fano -- Feshbach
resonance~\cite{HoubiersSF,Houbiers12,OHaraScience}, where the
scattering length for zero-energy s-wave collisions is large
compared to the interparticle spacing.

In a strongly-interacting 50-50 mixture of spin-up and spin-down
atoms, evidence of new, high-temperature superfluidity has been
found in the studies of collective
dynamics~\cite{OHaraScience,Kinast,Bartenstein,KinastMagDep} and
confirmed by the observation of quantized vortex
lattices~\cite{KetterleVortices}. Microscopic properties of the
superfluid have been probed by detecting pairs of Fermi atoms in
projection experiments~\cite{Jincondpairs,Ketterlecondpairs},
radiofrequency spectroscopy experiments~\cite{GrimmGap,Jingap},
and optical spectroscopy measurements of the order
parameter~\cite{HuletSpectr}. Measurements of the heat
capacity~\cite{JointScience}, collective
damping~\cite{KinastDampTemp}, and entropy~\cite{LuoEntropy} near
resonance reveal transitions in behavior close to the temperature
predicted for the onset of Fermi superfluidity.  Despite the
microkelvin range of temperatures, the phenomenon is referred to
as ``high-temperature superfluidity'' because the transition
temperature is a large fraction ($\simeq 30\%$) of the Fermi
temperature.

Measurements of the physical properties of strongly interacting
Fermi atoms are not limited to atomic physics and can be extended
to other Fermi systems: Fermi gases near a broad Feshbach
resonance exhibit universal
interactions~\cite{OHaraScience,Heiselberg,MechStab} and universal
thermodynamics~\cite{HoUniversalThermo,DrummondUniversal}, i.~e.
the properties of the gas do not depend on the details of the
interaction potentials (as long as the potential range is small)
and are identical to those of other resonantly-interacting Fermi
systems. For example, in a uniform strongly interacting gas, the
ground-state energy is a universal fraction, denoted $1+\beta$, of
the energy of a noninteracting gas at the same
density~\cite{OHaraScience,MechStab}. The universal energy
relationship was originally explored theoretically in the context
of nuclear matter~\cite{Heiselberg,Baker} and has later been
measured using ultracold Fermi
atoms~\cite{OHaraScience,Grimmbeta,JosephSound,HuletPhaseSeparation,SalomonExpInt,JinMomentum}.

The strongly-interacting atomic gas behaves as a fluid over a wide
range of temperatures~\cite{KinastDampTemp,ThomasUniversal}. In
the lower temperature range, this hydrodynamic behavior is
explained by superfluidity~\cite{Kinast,KinastDampTemp}. At
temperatures well above the superfluid transition, the observed
hydrodynamic frequency and the measured damping rate are not
consistent with a model of a collisional normal
gas~\cite{KinastDampTemp,BruunViscous,BruunViscosNormalDamping}.
The microscopic mechanism that provides hydrodynamic properties at
high temperatures is an open question.

In this paper, we review the hydrodynamic properties of a
strongly-interacting gas consisting of an equal mixture of spin-up
and spin-down atoms. We also present model-independent data on
hydrodynamic oscillations of the gas, i.e., the frequency and
damping rate of the radial breathing mode~\cite{KinastDampTemp} as
a function of the energy. These results should be useful for
future comparison with theoretical models.

The search for a fluid with the viscosity at the quantum minimum
is another feature of our  study of strongly-interacting atomic
Fermi gases. On general grounds, Kovtun et al.~\cite{Kovtun},
predicted that for any fluid the ratio of shear viscosity $\eta$
to entropy density $s$ is always $\ge\hbar/4\pi k_B$ ($k_B$ is
Boltzmann's constant). For a resonantly-interacting atomic gas, we
estimate the ratio $\eta/s$ from previous measurements of
damping~\cite{KinastDampTemp} and entropy~\cite{LuoEntropy}. We
discuss whether the atomic gas can be used for testing the
fundamental $\eta/s$ limit.

\section{Experimental system}

The starting point of our experiments is a highly degenerate,
strongly-interacting Fermi gas of $^6$Li.  The cloud is prepared
using  evaporation of an optically-trapped, 50-50 mixture of
spin-up/down states at 840 G, just above the center of a broad
Feshbach
resonance~\cite{OHaraScience,Kinast,KinastMagDep,JointScience,KinastDampTemp}.
To reduce the temperature, we do not employ a magnetic sweep from
a molecular BEC, in contrast to several other
groups~\cite{Bartenstein,Ketterlecondpairs,SalomonBEC}. Instead,
we evaporate directly in the strongly attractive regime, nearly on
the Feshbach resonance~\cite{OHaraScience}.

During the forced evaporation, the depth of the CO$_2$ laser
optical trap is reduced to a small fraction ($0.002-0.0005$) of
its maximum value ($\simeq 700$ $\mu$K), and then recompressed to
the final value, usually 4.6\% of the maximum trap depth for the
breathing mode experiments~\cite{KinastDampTemp}. From the trap
frequencies measured at 4.6\% of full trap depth, and corrected
for anharmonicity, we obtain the trap aspect ratio $\lambda
=\omega_z/\omega_\perp=0.045$
($\omega_\perp=\sqrt{\omega_x\omega_y}$) and the mean oscillation
frequency
$\bar{\omega}=(\omega_x\omega_y\omega_z)^{1/3}=2\pi\times 589(5)$
Hz. The shape of the trap slightly departs from cylindrical
symmetry: $\omega_x/\omega_y=1.107(0.004)$. Typically, the total
number of atoms after cooling is $N=2.0(0.2)\times 10^5$. The
corresponding Fermi energy $E_F=k_BT_F$ and Fermi temperature
$T_F$ at the trap center for a noninteracting gas  are given by
$T_F=(3 N)^{1/3}\hbar\bar{\omega}/k_B\simeq 2.4\,\mu$K at 4.6\% of
the maximum trap depth. For these conditions, the coupling
parameter of the strongly-interacting gas at $B=840$ G is
$k_Fa\simeq -30.0$, where $\hbar k_F=\sqrt{2m\, k_B T_F}$ is the
Fermi momentum, and $a=a(B)$ is the zero-energy scattering length
estimated from the measurements of Bartenstein et
al.~\cite{BartensteinFeshbach}.

This completes preparation of the gas at nearly the ground state.
The energy of the gas is increased from the ground state value by
abruptly releasing the cloud and then recapturing it after a short
expansion time $t_{heat}$. During the expansion time, the total
kinetic and interaction energy  is conserved. When the trapping
potential $U(\mathbf{x})$ is reinstated, the potential energy of
the expanded gas is larger than that of the initially trapped gas,
increasing the total energy by a known factor~\cite{JointScience}.
After waiting for the cloud to reach equilibrium, the sample is
ready for subsequent measurements.  The energy is then directly
measured in model-independent way from the mean square axial cloud
size as described below.

\section{The thermodynamic parameter: Energy or Temperature}

Equilibrium thermodynamic properties of the trapped gas, as well
as dynamical properties, can be measured as functions of the
temperature $T$ or of the total energy $E_{tot}$. Knowledge of one
such variable completely determines the properties of a universal
system with a known particle number $N$ (provided the local
density approximation is valid).

 To date, two  model-independent thermometry methods
have been reported for a strongly-interacting gas. The thermometry
method of the MIT group~\cite{KetterleShapeTransition} is only
applicable to imbalanced mixtures of spin-up and spin-down atoms
and is based on fitting the non-interacting edge of the majority
cloud by a Thomas-Fermi distribution with unconstrained
temperature. The  method of the Duke group is based on measuring
the entropy as a function of energy,
$S_{tot}(E_{tot})$~\cite{LuoEntropy}. This method has been
reported for a balanced mixture of spin-up and spin-down fermions.
Noting that $1/T=\partial S_{tot}/\partial E_{tot}$, it has been
proposed that for known energy the temperature can be obtained by
differentiating a fit to the $S_{tot}(E_{tot})$ data. In this
method, however, it is necessary to parametrize the data.

The other two known thermometry methods are model dependent and
rely on magnetic field sweeps between the molecular BEC or
noninteracting gas regime and the strongly interacting
regime~\cite{GrimmGap,JinThermometry}. The temperature of the
strongly interacting gas is then estimated from that measured in
the BEC or noninteracting gas using a theoretical model of the
entropy~\cite{QijinThermo}. The other thermometry method of the
Duke group~\cite{JointScience} is based on comparing the measured
density distribution with an approximate model for the density
profiles~\cite{LevinDensity}.

A model-independent parametrization of the data is provided by
measuring the energy of the trapped gas. In the universal regime,
the energy per particle $E=E_{tot}/N$ can be easily measured
because the trapped gas obeys the virial
theorem~\cite{ThomasUniversal}:
\begin{equation}
\langle U(\mathbf{x})\rangle=\frac{E}{2},\label{eq:virial}
\end{equation}
where $U(\mathbf{x})$ is the harmonic trapping potential and
$\langle U(\mathbf{x})\rangle$ represents average potential energy
per particle. This result is remarkable: Despite the complicated
many-body strongly interacting ground state, the gas obeys the
same virial theorem as a non-interacting Fermi gas. This theorem
provides a simple  model-independent  energy measurement, just
from the size of the cloud. In the local density approximation,
with an isotropic pressure, it is easy to show from
Eq.~\ref{eq:virial} that the energy per particle in a 50-50
mixture of two spin states in a harmonic trap is
\begin{equation}
E = 3m\omega_z^2\langle z^2\rangle , \label{eq:energy}
\end{equation}
where $z$ is the axial direction of the cigar-shaped cloud. The
energy can also be measured from the radial dimension of the
expanded cloud in the hydrodynamic regime, where the expansion
factor is known~\cite{OHaraScience}.

For the radial breathing mode near the Feshbach resonance, we
present both the frequency and damping rate data as a function of
energy per particle to provide  model-independent results.

\section{Hydrodynamic flow of a strongly interacting Fermi gas}

Hydrodynamic flow was observed in the very first and simplest
experiments with a strongly-interacting Fermi gas, where the gas
was released from a cigar-shape trap~\cite{OHaraScience}. The
signature of hydrodynamic flow was the anisotropic expansion of
the cigar cloud into a disc, i.e. the flow was in the direction of
the largest pressure gradient, inverting the aspect ratio after
expansion.

In-trap hydrodynamics can be studied by observing the compression
(breathing) mode of the
gas~\cite{Kinast,Bartenstein,KinastMagDep,KinastDampTemp,GrimmLeeHuangYang}.
There are two distinct frequency regimes, for a hydrodynamic fluid
and for a collisionless normal gas. In a cigar-shape trap with our
parameters, for the radial mode, these frequencies are
$\omega=1.84\,\omega_\perp$ and $\omega=2.10\,\omega_\perp$,
respectively. In the universal gas, the hydrodynamic frequency has
been predicted to stay the same at any temperature as long as the
flow remains isentropic~\cite{ThomasUniversal}.

\begin{figure}[htb]
\centerline{\includegraphics[width=4.0in,clip]{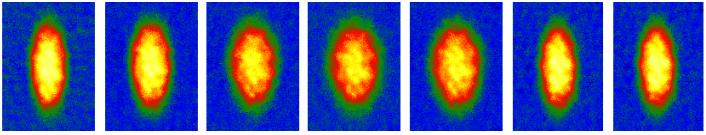}}
\caption{Snap-shots of gas after oscillation for a variable time
$t_{hold}$, followed by release and expansion for 1 ms. Each
measurement is destructive. $t_{hold}$ is increasing from left to
right.} \label{fig:SnapShots}
\end{figure}

\begin{figure}[htb]
\centerline{\includegraphics[width=3.0in,clip]{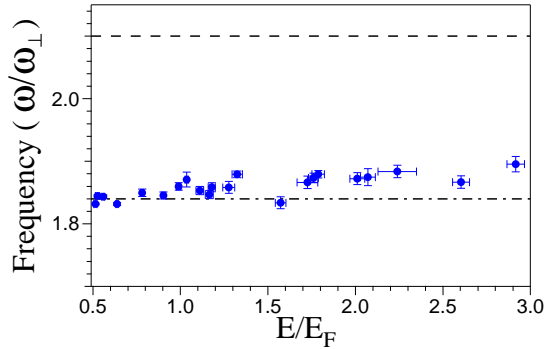}}
\caption{Breathing mode in a strongly-interacting gas. Normalized
frequency versus the normalized energy per particle. The
superfluid phase transition is located at $E_c\simeq E_F$. The
upper dotted line shows the frequency for a noninteracting gas.}
\label{fig:BreathingMode}
\end{figure}

\begin{figure}[htb]
\centerline{\includegraphics[width=3.0in,clip]{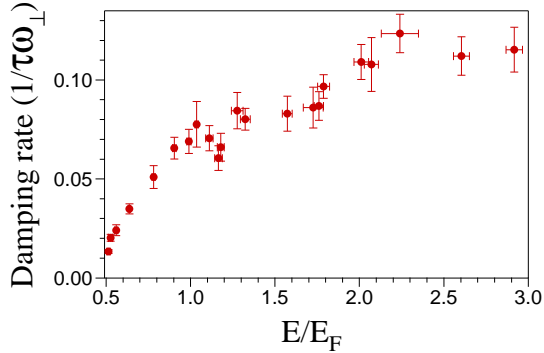}}
\caption{Breathing mode in a strongly-interacting gas. Normalized
damping versus the normalized energy per particle.}
\label{fig:BreathingModeDamping}
\end{figure}

In the experiment, the radial breathing mode is excited by
releasing the cloud and recapturing the atoms after 25 $\mu$s (for
4.6\% of the maximum trap depth). After the excitation, we let the
cloud oscillate for a variable time $t_{hold}$, at the end of
which the gas is released and imaged after $\simeq 1$ ms of
expansion~\cite{Kinast}. The snap-shots of the oscillating gas are
shown in Fig.~\ref{fig:SnapShots}.
The measured frequency (Fig.~\ref{fig:BreathingMode}) remains at
the hydrodynamic value over the nominal temperature range
$T=0.12-1.1\,T_F$~\cite{KinastDampTemp}, which corresponds to an
energy range from nearly the ground state value $\simeq 0.5\,E_F$
to $3.0\,E_F$.
The frequency stays far below the frequency of a non-interacting
gas. No signatures of the superfluid transition are seen in the
frequency dependence, but we can obtain several estimates of the
critical energy from measurements of other thermodynamic
quantities, such as the entropy and heat capacity. In these two
quantities, a change in behavior is observed at $E_c=0.94\,E_F$
($T=0.29\,T_F$) and $E_c=0.85\,E_F$ ($T=0.27\,T_F$), respectively.
In Figure~\ref{fig:BreathingModeDamping}, we display the
normalized damping rate $1/\omega_\perp\,\tau$ versus the energy.
In contrast to the frequency, the damping rate exhibits
interesting features. At $E_c=1.01\,E_F$ one observes a clear
change in the damping versus energy dependance: The monotonic rise
switches to flat dependance, which might be a signature of a phase
transition.

In examining these features, it is appropriate to ask: What are
the microscopic mechanisms that cause the gas to have hydrodynamic
properties? Hydrodynamic behavior can appear via at least two
mechanisms: (i) superfluidity and (ii) normal dynamics of atomic
gas with large number of collisions. Below $E_c=1.01\,E_F$, the
reduction in the  damping rate as $T\rightarrow 0$ is consistent
with superfluidity, and inconsistent with the scenario of a
collisional normal gas~\cite{Kinast,KinastDampTemp}.  Above $E_c$,
however,  collisional dynamics of a normal atomic gas does not
completely explain the observed behavior. For such a normal
system, Bruun and Smith~\cite{BruunViscosNormalDamping} found that
hydrodynamic behavior is only possible at $T< 1.4\,T_F$, i.e. only
for  temperatures where the collision rate is large enough. The
highest observed damping rate$1/\tau=0.12\,\omega_\perp$ occurs
for $E>2.2\,E_F$ and is consistent with predictions for a
transition from hydrodynamic to collisionless
behavior~\cite{BruunViscosNormalDamping}. However, the high
damping rate is predicted to occur with a higher frequency, closer
to $2.10\,\omega_\perp$~\cite{BruunViscosNormalDamping}.

It is possible that above the phase transition, the dynamics is
significantly affected by the presence of non-condensed pairs. The
step in damping rate at about $E\simeq 2E_F$ might be due to
pair-breaking since at this energy, the trap-averaged binding
energy becomes smaller than $\hbar\omega$~\cite{KinastDampTemp}. A
pseudogap formalism~\cite{JointScience,LevinDensity} as well as
the recent observation of the MIT
group~\cite{KetterlePairingWOutSF} suggests that atoms can be
paired at temperatures well above the phase transition.

\section{Quantum viscosity}

In a  strongly interacting system Fermi gas, where the
interparticle separation $l\propto 1/k_F$ sets the length scale,
there is a natural unit of shear viscosity $\eta$ which has
dimension of momentum divided by cross section. The relevant
momentum is the Fermi momentum, $\hbar k_F=\hbar /l$. The relevant
area is determined by the unitarity-limited collision cross
section $4\pi/k_F^2\propto l^2$. Hence, $\eta \propto \hbar /l^3
=\hbar\,n$, where $n$ is the local total density, so that
\begin{equation}
\eta = \alpha\, \hbar\, n. \label{eq:qmviscosity}
\end{equation}
Here, $\alpha$ is generally a dimensionless function of the local
reduced temperature $T/T_F(n)$, where $T_F(n)\equiv
\hbar^2(3\pi^2n)^{2/3}/2mk_B$ is the local Fermi temperature.
Eq.~\ref{eq:qmviscosity} has been discussed by
Shuryak~\cite{Shuryak}. We see that viscosity has a natural
quantum scale, $\hbar\,n$. If the coefficient $\alpha$ is of order
unity or smaller, we say that the system is in the quantum
viscosity regime.

Using string theory methods, Kovtun et al., have shown that for a
wide class of strongly interacting quantum fields, the ratio of
the shear viscosity to the entropy density has a universal minimum
value~\cite{Kovtun}. The entropy density $s$ has units of
$n\,k_B$. Hence, in the ratio $\eta /s$ the density cancels so
that $\eta /s$ has natural units of $\hbar /k_B$. The string
theory prediction is
\begin{equation}
\frac{\eta}{s}\geq\frac{1}{4\pi}\frac{\hbar}{k_B}.
\label{eq:minviscosity}
\end{equation}

An important question is then how close  a strongly interacting
Fermi gas comes to the minimum quantum viscosity limit. Answering
this question requires the determination of two physical
quantities, shear viscosity and entropy density. To estimate the
ratio $\eta/s$, we separately integrate the numerator and
denominator over the trap volume. Then using $\int d^3\mathbf{x}\,
n =N$, where $N$ is the total number of atoms and $\int
d^3\mathbf{x}\, s =S_{tot}$ is the total entropy, one obtains,
\begin{equation}
\frac{\eta}{s}\simeq\frac{\int d^3\mathbf{x}\, \eta}{\int
d^3\mathbf{x}\, s}= \frac{\hbar}{k_B}\frac{\langle
\alpha\rangle}{S/k_B}. \label{eq:viscosityoverentropy}
\end{equation}
Here, $\langle \alpha\rangle\equiv(1/N)\int
d^3\mathbf{x}\,n\,\alpha(\mathbf{x})$ is the trap average of the
dimensionless universal function $\alpha$ that determines the
shear viscosity according to Eq.~\ref{eq:qmviscosity} and
$S=S_{tot}/N$ is the entropy per particle which has been measured
as a function of energy~\cite{LuoEntropy}.

We estimate $\langle\alpha\rangle(E)$  from the damping rate
$1/\tau$ of the radial breathing mode,
Fig.~\ref{fig:BreathingModeDamping}.   {\it If} the damping rate
arises from shear viscosity, then the value of $\langle
\alpha\rangle$ is readily determined. Using the shear pressure
tensor~\cite{LandauFluids} in the hydrodynamic equations of motion
for the radial breathing mode~\cite{ThomasUniversal}, we easily
obtain
\begin{equation}
\frac{1}{\tau\omega_\perp}= \frac{\hbar\omega_\perp}{E}\langle
\alpha\rangle. \label{eq:damping}
\end{equation}

\begin{figure}[htb]
\centerline{\includegraphics[width=3.0in,clip]{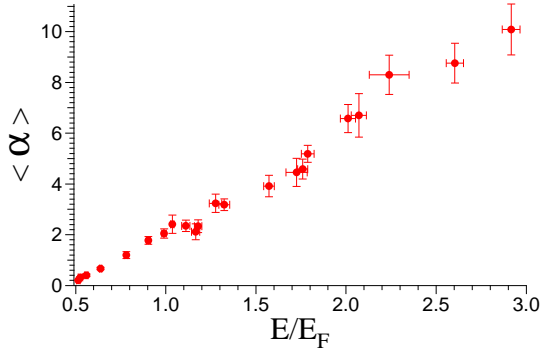}}
\caption{Quantum viscosity in strongly-interacting Fermi gas. The
local shear viscosity takes the form $\eta = \alpha\,\hbar\,n$. In
the figure, $\langle\alpha\rangle$ is a trap-averaged value of the
dimensionless parameter $\alpha$. } \label{fig:alpha}
\end{figure}

We determine the energy per particle in a model independent way by
exploiting the virial theorem, as discussed
above~\cite{LuoEntropy,ThomasUniversal}. Using our measured
damping ratios $1/(\tau\omega_\perp)$  as a function of energy, we
determine $\langle\alpha\rangle (E)$, as shown in
Fig.~\ref{fig:alpha}.

\begin{figure}[htb]
\begin{center}
\includegraphics[height=4.0in]{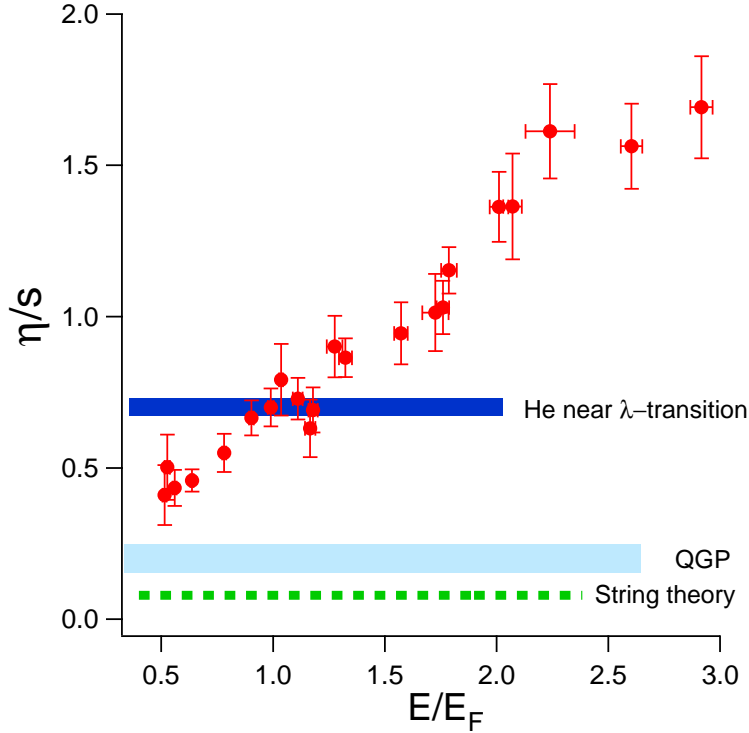}
\caption{Ratio of the shear viscosity $\eta$ to the entropy
density $s$ for a strongly interacting Fermi gas as a function of
energy $E$, red solid circles. The lower green dotted line shows
the string theory prediction $1/(4\pi)$. The light blue bar shows
the estimate for a quark-gluon plasma (QGP)~\cite{BassPrivate},
while the blue solid bar shows the estimate for $^3$He and $^4$He,
near the $\lambda$- point. } \label{fig:minviscosity}
\end{center}
\end{figure}

Using the values of $\langle\alpha\rangle(E)$ from
Fig.~\ref{fig:alpha} and the measured entropy $S(E)$ from
Ref.~\cite{LuoEntropy} in Eq.~\ref{eq:viscosityoverentropy}, we
estimate the ratio of the viscosity to entropy density, as shown
in Fig.~\ref{fig:minviscosity}. A similar plot has been given
previously by Thomas Sch\"{a}fer, based on our damping and entropy
measurements~\cite{SchaferRatio}. In the plot, note that
$E_0\simeq 0.5\,E_F$ is the ground state energy and $E_F$ is the
Fermi energy of a noninteracting gas at the trap center. The
superfluid transition occurs near $E=E_F$, above which the gas is
normal. For comparison, the estimated value for $^3$He and $^4$He
near the $\lambda$-point is $\eta/s\simeq 0.7$. For a quark-gluon
plasma, a current theoretical estimate~\cite{BassPrivate} is
$\eta/s = 0.16-0.24$.

While our initial estimates for the $\eta/s$ ratio place the
strongly interacting Fermi gas in the quantum viscosity regime, a
caveat for our result is that we have not proven that the damping
arises from viscosity~\cite{KinastDampTemp}. Other sources of
damping may contribute, for example Landau
damping~\cite{ShlyapnikovLandauDamping}. Thus, the actual viscosity may be lower. Our data
may not be determining the minimum viscosity, which may make a
contribution that is smaller than the total measured damping rate.

\section{Conclusion}

We have explored the fluid properties of a strongly interacting
Fermi gas. Measurements of the frequency and damping of the radial
breathing mode as a function of energy provide model-independent
data that indicate very low viscosity, fluid-like behavior over a
wide range of energies covering both the superfluid and normal
phases. The measured frequency agrees with the macroscopic model
of universal isentropic hydrodynamics. The unitary collision
dynamics that provides this fluid-like behavior well above the
phase transition is not completely understood. Our estimate of the
viscosity over entropy density ratio ($\eta/s$) places strongly
interacting Fermi gases in the quantum viscosity regime.

\section*{Acknowledgements}
This research has been supported by the Army Research Office and
the National Science Foundation, the Physics for Exploration
program of the National Aeronautics and Space Administration, and
the Chemical Sciences, Geosciences and Biosciences Division of the
Office of Basic Energy Sciences, Office of Science, U. S.
Department of Energy.


\end{document}